\documentclass[prb,11pt]{revtex4-1}

\usepackage{amsmath}    
\usepackage{graphicx}   
\usepackage{verbatim}   
\usepackage{color}      
\usepackage{subfigure}  
\usepackage{hyperref}   
\usepackage{latexsym}
\usepackage[normalem]{ulem}

\begin{document}

\title{Disentangling defect-induced magnetism in SiC}

\author{Yutian Wang $^{1,4}$}
\author{Lin Li $^{1,2}$}
\author{Slawomir Prucnal$^1$}
\author{Xuliang Chen$^3$}
\author{Wei Tong$^5$}
\author{Zhaorong Yang$^3$}
\author{Frans Munnik$^1$}
\author{Kay Potzger$^1$}
\author{Wolfgang Skorupa$^1$}
\author{Sibylle Gemming$^1$}
\author{Manfred Helm$^{1,4}$}
\author{Shengqiang Zhou$^1$}
\email[E-mail: ]{s.zhou@hzdr.de}

\affiliation{1. Institute of Ion Beam Physics and Materials Research, Helmholtz-Zentrum Dresden-Rossendorf(HZDR),P.O.Box 510119,01314 Dresden,Germany}
\affiliation{2. Department of Physics and Electronics, School of Science, Beijing University of Chemical Technology, Beijing 100029, China}
\affiliation{3. Key Laboratory of Materials Physics, Institute of Solid State Physics, Chinese Academy of Sciences, Hefei 230031, People's Republic of China}
\affiliation{4. Technische Universit\"at Dresden, 01062 Dresden, Germany }
\affiliation{5. High Magnetic Field Laboratory, Hefei Institutes of Physical Science, Chinese Academy of Sciences, Hefei 230031, People's Republic of China}

\begin{abstract}
We present a detailed investigation of the magnetic properties in SiC single crystals bombarded with Neon ions. Through careful measuring the magnetization of virgin and irradiated SiC, we decompose the magnetization of SiC into paramagnetic, superparamagnetic and ferromagnetic contributions. The ferromagnetic contribution persists well above room temperature and exhibits a pronounced magnetic anisotropy. We qualitatively explain the magnetic properties as a result of the intrinsic clustering tendency of defects.
\end{abstract}

\maketitle
\section{Introduction}
Defect-induced ferromagnetism, also referred as ``$d^0$ ferromagnetism'' in
contrast to traditional ferromagnets containing partially filled 3\textit{d}
or 4\textit{f} electrons, is continuously attracting research interest\cite{Coey201012,Coey2005660}.
It not only challenges the basic understanding of ferromagnetism, but also
allows for potential spintronic applications, such as magneto-optics
components\cite{PhysRevLett.96.197208}. Defect-induced ferromagnetism was first observed in carbon-based materials\cite{PhysRevB.66.024429,PhysRevLett.91.227201,ADMA:ADMA200305194}. Ferromagnetism was subsequently
 reproduced in graphite by different groups\cite{PhysRevLett.95.097201,ISI:000262292200007,ISI:000271895500020,Yang20091399,PhysRevB.83.085417,He20111931,Shukla20121817} and also in other carbon-based materials\cite{Makarova20031575,han2003observation,PhysRevB.72.224424,PhysRevB.75.075426,ISI:000302630100016}. Later on, defect-induced
  ferromagnetism was observed in various
oxide insulators, including HfO$_2$\cite{ISI:000223085400033},
TiO$_2$\cite{0953-8984-18-27-L01,PhysRevB.79.113201},
ZnO\cite{PhysRevLett.99.127201,PhysRevB.80.035331,PhysRevLett.104.137201,0022-3727-41-10-105011}
and very recently in MoS$_2$\cite{Mathew2012101,Tongay2012101}.
However, the necessary use of thick substrates for growing those
oxides makes the interpretation of magnetometry data complicated,
\emph{i.e.} one has to exclude any ferromagnetic contamination in
their substrates\cite{PhysRevB.81.214414,ISI:000267045400001}. For
carbon-based material, Ohldag \textit{et al}., have shown that the
magnetic order originates from the carbon $\pi$-electron system in
proton irradiated graphite, or from hydrogen-mediated electronic
states in untreated
graphite\cite{PhysRevLett.98.187204,1367-2630-12-12-123012}. On
the other hand, the authors confirmed that the main part of the
magnetic order exists in the near-surface region
\cite{1367-2630-12-12-123012}, making the total measured moments
smaller (around 1${_....}$ $10\times10^{-6}$ emu per
sample)\cite{PhysRevLett.98.187204}.

Very recently, SiC single crystals have been shown to be ferromagnetic after
ion and neutron bombardment\citep{Li201198,PhysRevLett.106.087205} or after Aluminum doping\cite{ISI:000264791800021}.
Tentatively, it is proposed that V$_{Si}$V$_C$ divacancies form local
magnetic moments arising from \textit{sp} states which couple ferromagnetically due
to the extended tails of the defect wave functions\cite{PhysRevLett.106.087205}. As an important
technologic material, SiC is commercially available at large scale and with
high quality at the microelectronic grade\cite{ISI:000223514900038,Matsunami20062}.
The impurity concentration of Fe or Ni is of the range of $10^{14}$/cm$^3$, which corresponds to 0.015 $\mu$g/g \cite{ISI:000182770100028}. This impurity level is around one order of magnitude lower than that of the purest graphite samples\cite{Esquinazi20101156}, for which the origin of the defect induced magnetism is still under intensive discussion\cite{0295-5075-97-4-47001,0295-5075-98-5-57006,Venkatesan2013279}. Therefore, SiC presents a suitable model system for studying defect-induced ferromagnetism in addition to carbon based materials.

In this paper, we present a systematic magnetic investigation of SiC prepared by Ne ion irradiation.
The possible Fe, Co or Ni contamination in SiC is excluded by particle induced X-ray emission (PIXE)
detection. By carefully measuring the magnetization of virgin and irradiated SiC, the induced
magnetization in SiC can be decomposed into paramagnetic, superparamagnetic and ferromagnetic
contributions. The ferromagnetic contribution persists well above room temperature. Those
results could promote further understanding of defect-induced ferromagnetism.

\section{Experimental methods}

To examine the impurities in the virgin wafer, we applied the PIXE
technique using 2 MeV protons with a broad beam of 1 mm$^2$. As
stated in Ref.~\onlinecite{Esquinazi20101156}, PIXE is accurate enough to detect trace impurities
in bulk volume that can lead to ferromagnetism larger than that
induced by the irradiation procedure. We used this technique to
examine SiC wafers from different suppliers (Cree and KMT
Corporation). Within the detection limit, transition metal
impurities (Fe, Co and Ni) can be excluded from all examined
wafers. In this study, we present the result based on one 6H-SiC
wafer purchased from KMT Corporation. The same wafer was cut into
smaller pieces of 10$ \times$10 mm$^2$ for further experiments.
Therefore, the starting materials in this report are well
controlled and free of variations. During the irradiation experiment, the samples were pasted on a Si wafer by carbon tapes to avoid the usage of metal clamps. After irradiation, the samples were carefully cleaned by rising with Acetone, Isopropanol and deionized water. During the whole procedure (cutting, irradiation, measurement), special cautions, such as only using ceramic or plastic tools, were paid to avoid metal contamination. Moreover, in order to exclude possible contamination during implantation, we applied Auger electron spectroscopy, which is a surface sensitive technique, to measure the samples at both the front and the back sides of the irradiated samples. For most elements, the detection limit of AES is in the range of 0.01-0.1 atom\%, which is around 10$^{12}$--10$^{13}$ atoms/cm$^{2}$ by assuming a detection depth of 2 nm. Within the detection limit, contamination by Fe, Co or Ni can be excluded. For more details about the PIXE and AES results, see ref.~\onlinecite{wang2014JAP}.

\begin{figure} \center
\includegraphics[scale=0.3]{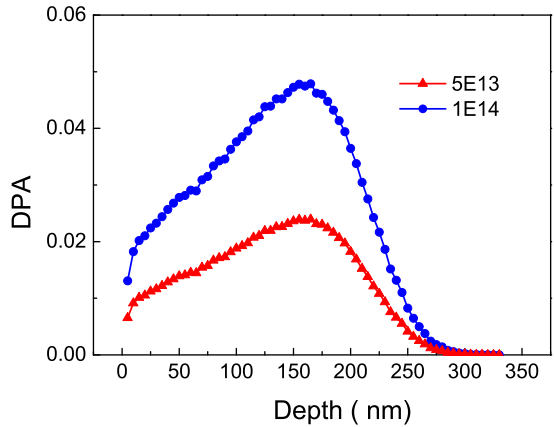}
\caption{SRIM simulation of displacement per atom (dpa) in SiC by 140 keV Ne ion irradiation. The labels indicate the ion fluence in cm$^{-2}$.} \label{fig1}
\end{figure}

The 10$\times$10 mm$^2$ SiC specimens with thickness of 0.5 mm
were implanted with Ne ions at an energy of 140 keV. The
implantation was done at room temperature without heating or
cooling the sample holder. The details of the implantation and
the magnetization dependence on the ion fluence were reported in Ref.~\onlinecite{Li201198}. In this paper, we select two samples with large
magnetization. The implantation fluences were 5$\times$10$^{13}$
cm$^{-2}$ to 1$\times$10$^{14}$ cm$^{-2}$, thereafter the samples
are referred to 5E13 and 1E14, respectively. The implantation
induced maximum displacement per atom (dpa) was calculated using
SRIM2003\cite{srim} and is shown in Fig. \ref{fig1}. The dpa
numbers have been estimated by using a mass density of 3.21
g/cm$^3$ and a displacement energy of 25 eV for both Si and C. In
SRIM calculation the heating effect was not taken into account.
Magnetometry was performed using a SQUID-MPMS or SQUID-VSM magnetometer from
Quantum Design. The electron spin resonance (ESR) spectroscopy was performed at 9.4 GHz using a Bruker spectrometer (Bruker ELEXSYS E500).  

\section{Results}

\subsection{Paramagnetism in virgin SiC}

\begin{figure}[!htbp]

\centering
\includegraphics[scale=0.3]{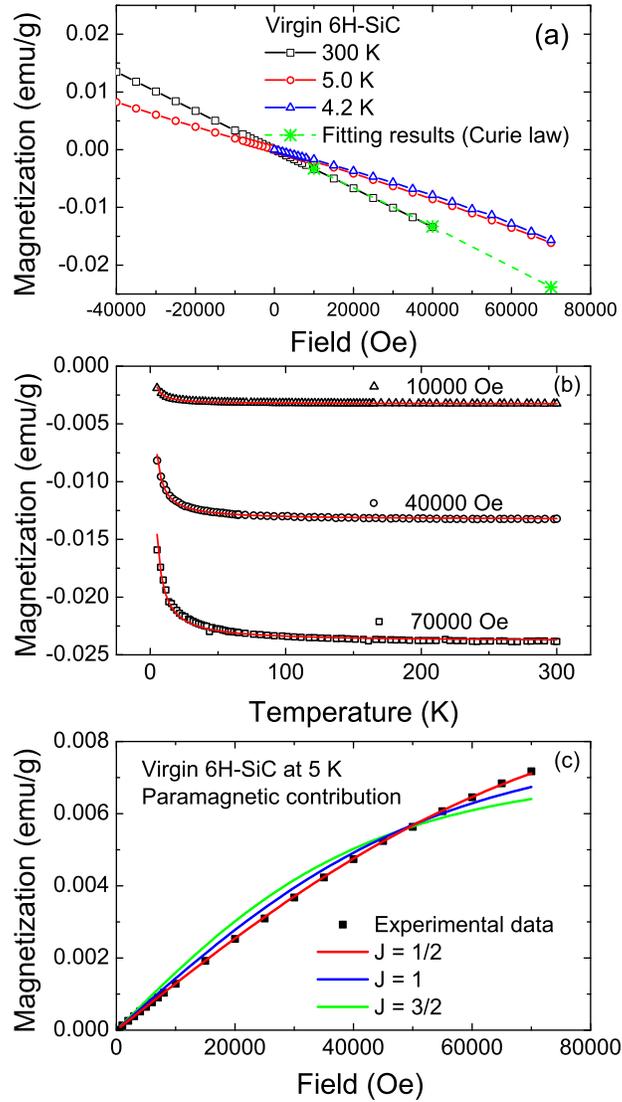}
\caption{(a) Magnetization of virgin 6H-SiC at different temperatures. At larger field and low temperature, one can see a deviation from the linear behavior: the indication of the presence of paramagnetism. (b) Temperature dependent magnetization (open symbols) of virgin 6H-SiC measured at different applied field. All curves can be fitted well (red solid lines) by considering a paramagnetic contribution with a diamagnetic background. (c) Fitting of the paramagnetism by Brillouin function with different \textit{J}. Tthe magnetization is normalized by the whole sample mass.} \label{fig2}
\end{figure}

\begin{figure} \center
\includegraphics[scale=0.3]{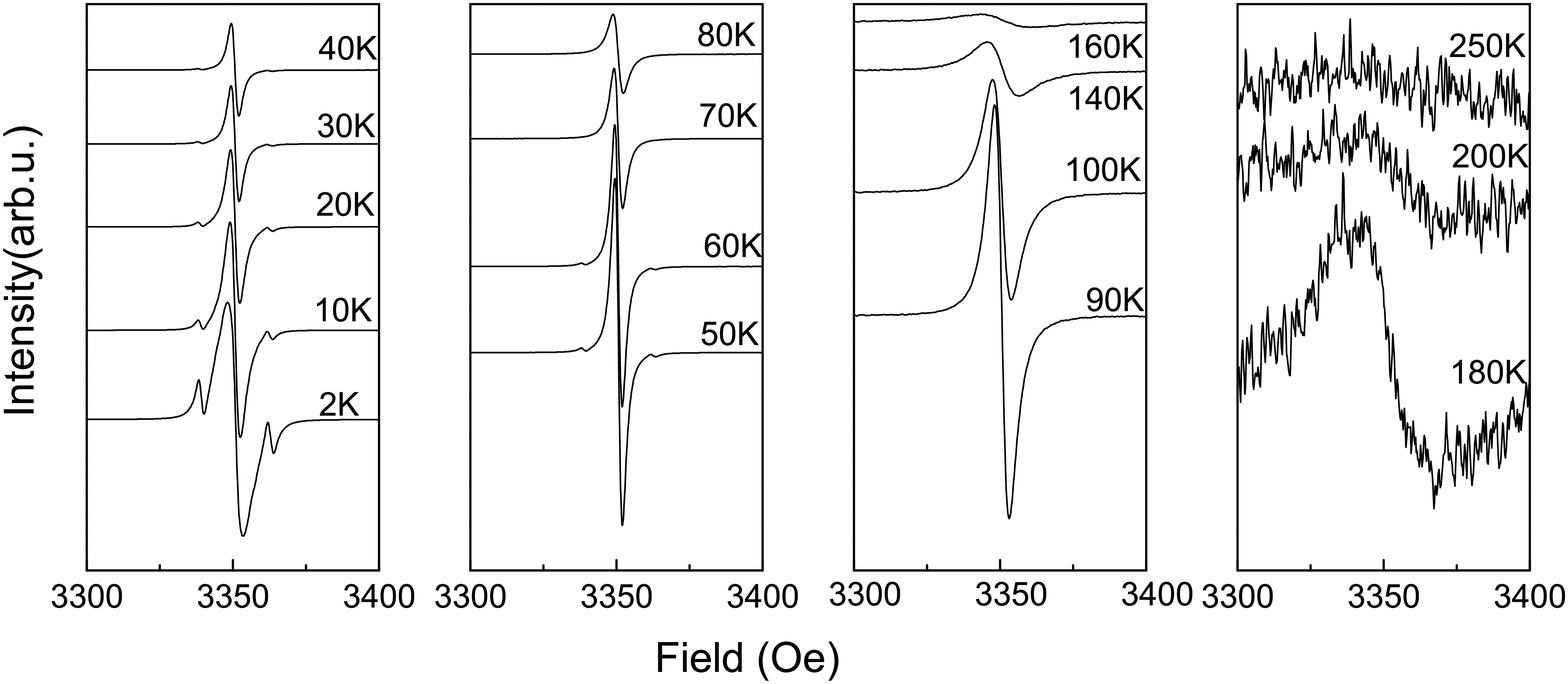}
\caption{ESR spectra for a virgin 6H-SiC sample used in this study at different temperature. The field is applied parallel to the sample surface [(0001) plane].} \label{ESR}
\end{figure}

The field-dependent magnetization of virgin 6H-SiC at 4.2 K, 5 K and 300 K
is shown in Fig. \ref{fig2}(a). No hysteresis loop is observed at any measured
temperature. The measurement at 300 K displays a typical diamagnetic
behavior. However, one can observe a clear deviation from the diamagnetism
at 4.2 K or 5 K. The curves are not linear, which can only be explained by
the presence of paramagnetism. Fig. \ref{fig2}(b) shows the temperature dependent
magnetization measured under different fields. For all curves, the large
contribution represents the diamagnetic background, which is essentially
temperature independent. We fit all the curves by Curie\textquoteright s
law:

\begin{align}
M=\chi_d+C\frac{B}{T},
\end{align}

where \textit{$\chi_d$} is the diamagnetic background, $B$ is the magnetic field,
$T$ is the temperature and $C$ is the Curie constant. The fitted diamagnetism
(\textit{$\chi_d$}) at different fields has been plotted in Fig. \ref{fig2}(a) by the
star symbols. The three values (at 10000, 40000, and 70000 Oe) fall on the line of the field dependent
magnetization measured at 300 K. Therefore, as a first order of
approximation the paramagnetic component can be extracted by subtracting the
diamagnetism measured at 300 K, as shown in Fig. \ref{fig2}(c). The paramagnetic part
can be well fitted by Brillouin function:

\begin{align}
M(x)=NJ{\mu_B}g{_J}[\dfrac{2J+1}{2J}coth(\dfrac{2J+1}{2J}{x})-\dfrac{1}{2J}coth(\dfrac{1}{2J}{x})],
\end{align}

where, the $g_J$ factor is about 2 obtained from electron spin
resonance measurement shown in Fig. \ref{ESR}, $\mu_B$ is the Bohr magneton, and N
is the density of spins. The curve can be well fitted by J = 0.5
corresponding to single electrons and $\textit{N}$ =
1.04$\times$10$^{18}$ $\mu_B$/g. Thus in the virgin sample, the
paramagnetic part originates from the electron spins, which
probably reside on defect sites. Previous investigations have
shown that isolated V$_{Si}$ and V$_C$ vacancies are the dominant
intrinsic paramagnetic defects in SiC\cite{ISI:000182770100028}. In high purity semi-insulating (HPSI) 4H-SiC\cite{ISI:000182770100028}, the intrinsic impurity concentration is of the order of 2$\times$10$^{16}$/cm$^3$. The impurities can induce Si or C vacancies and anti-sites. So the defect concentration can be estimated to be 1$\times$10$^{17}$/cm$^3$. In HPSI 6H-SiC \cite{Hobgood1995}, the impurity concentration is generally higher by one order of magnitude than in 4H-SiC. Therefore, it is not surprising to have a spin density of 3.2$\times$10$^{18}$ $\mu_B$/cm$^3$ related to defects and impurities. 

Paramagnetic centers in the virgin SiC have also been detected by ESR. The ESR spectra exhibit a central line and some hyperfine lines. The \emph{g} factor is calculated to be around 2.005, which is the characteristic of free electrons. From the line shape, the defect is very probably a negatively charged Si vacancy \cite{PhysRevB.68.165206}. The definitive identification requires a detailed angular scan to see the anisotropy of the hyperfine lines and a large amount of defects by neutron or electron irradiation \cite{PhysRevB.68.165206}, which is beyond the scope of this work. It is worthy to note that $S=1/2$ spins do not exhibit magnetocrystalline anisotropy as the resonance field of the central line does not change with the sample orientation \cite{PhysRevB.68.165206}.

\subsection{Magnetic phases in Ne irradiated SiC}
\begin{figure}[!htbp]
\centering
\includegraphics[scale=0.3]{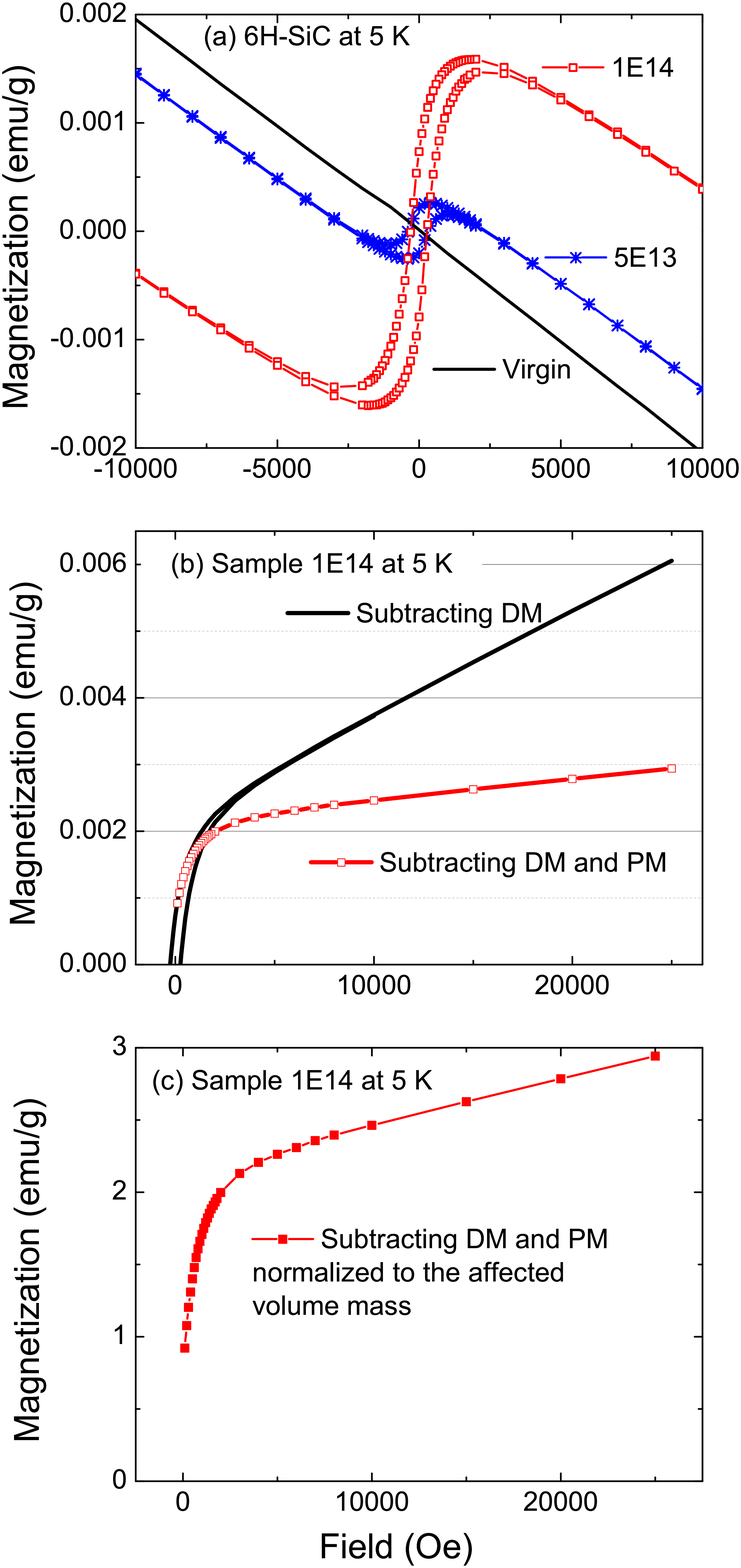}
\caption{(a) Comparison of the hysteresis loops measured at 5 K for the virgin SiC and samples 1E14 and 5E13. (b) and (c) Magnetization at 5 K of sample 1E14 after subtracting different contributions from the substrate. DM: diamagnetism and PM: paramagnetism. Before and after subtracting the diamagnetic background from the substrate, the magnetization is normalized by the whole sample mass (a, b) and by the implanted mass (c), respectively. The implanted depth corresponds to around 0.1\% of the sample thickness.}\label{fig4}
\end{figure}

In order to get a more precise picture of the magnetic properties
in the Ne irradiated samples, we have to consider the paramagnetic
contribution in the virgin SiC, since most of the volume remains
intact. Figure \ref{fig4}(a) shows the comparison of the
hysteresis loops measured at 5 K for the virgin SiC and for samples 
1E14 and 5E13. In Fig. \ref{fig4}(b), we show the magnetization of sample
1E14 after subtracting the diamagnetism and the paramagnetism
originating from the substrate. After this data treatment, the
paramagnetic part can be mostly subtracted, but not all. The
slightly increased magnetization at the large-field part (shown in
Fig. \ref{fig4}(b) and (c)) indicates more than one magnetic contribution
in the irradiated sample. Nevertheless, we can estimate roughly the saturation magnetization to be around 0.3--0.4 $\mu_B$ per vacancy calculated from the SRIM simulation (see Fig. \ref{fig1}). In reality, this value can be larger, as we possibly overestimate the amount of vacancies by neglecting the dynamic annealing effect in the simulation. 

\begin{figure}
\centering
\includegraphics[scale=0.3]{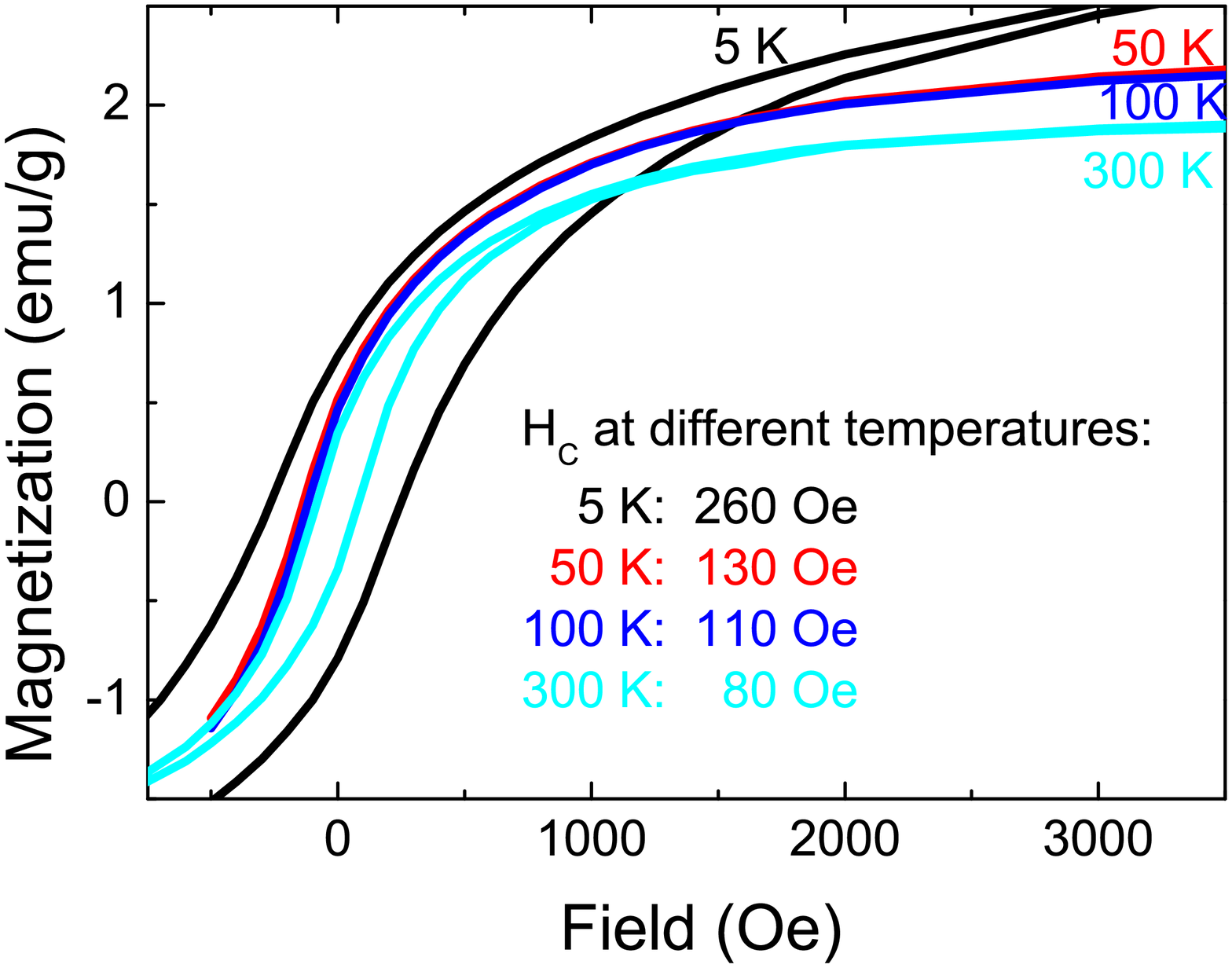}
\caption{Hysteresis loops measured at different temperatures for
sample 1E14. One can see a fast drop of the coercivity from 5K to
50K, but a slower decrease above 50 K. The magnetization is normalized by the implanted mass corresponding to around 0.1\% of the sample.}\label{fig5}
\end{figure}

Figure \ref{fig5} shows the hysteresis
loops of sample 1E14 measured at different temperatures. One can
observe three features: (1) the hysteresis behavior persists above
300 K, (2) there is only a slight decrease of the saturation
magnetization, and (3) the coercivity changes for different
temperatures. The coercivity drops from 260 Oe at 5 K to 130 Oe at
50 K, and then drops slowly with increasing temperatures. These
features indicate the co-existence of ferromagnetism and superparamagnetism \cite{beam1959,Shiratsuchi2004141,PhysRevB.80.094409}. 

\begin{figure}
\includegraphics[scale=0.2]{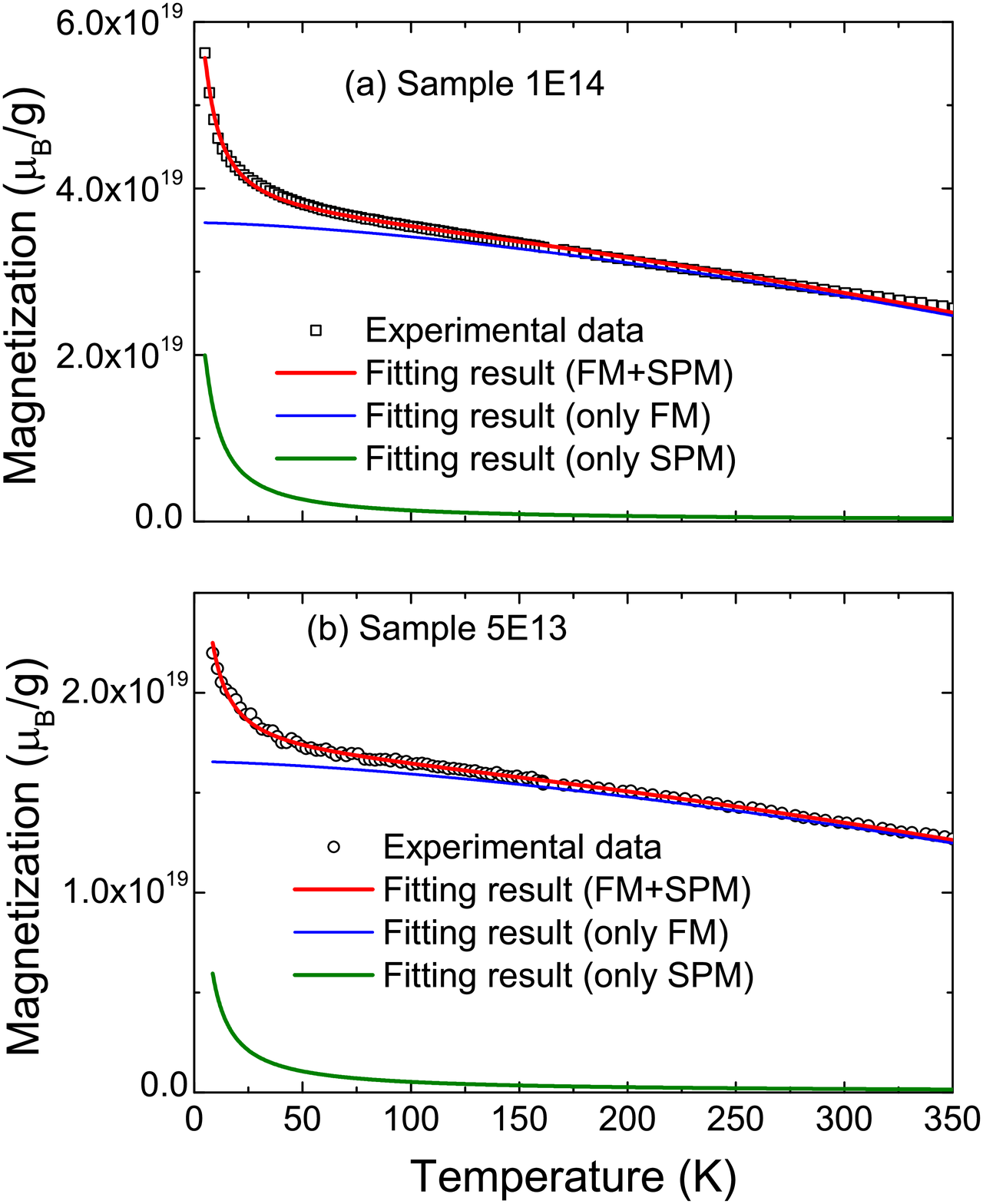}
\caption{Magnetic remanence of Ne irradiated SiC samples: one can
see the large difference of the magnetization vs. temperature at
temperature below and above 50 K. The data are shown without correcting for the diamagnetic and paramagnetic background which is negligible due to the small field. The solid lines are fits according to Eq. 3.}\label{fig6}
\end{figure}

Now we have to find an approach to separate the superparamagnetic
and the ferromagnetic contributions in the Ne irradiated SiC.
While it is difficult to calculate separate hysteresis loops for
both contributions, the magnetic remanence and coercivity for
superparamagnetism show distinct behavior compared with
ferromagnetism upon increasing temperature. Therefore, we can
analyze the temperature dependent magnetic remanence as shown in
Fig. \ref{fig6} in order to decompose it into different
components. For superparamagnetism, the remanence and coercivity
are more sensitive to the temperature increase \cite{Shiratsuchi2004141}. To measure the remanence, we
first cooled the sample to 5 K at a large field (10000 Oe). Then the
field was set to 10 Oe to compensate a possible small negative field
in the superconducting magnet. The remanence was measured during
warming up. As shown in Fig. \ref{fig6}, the magnetic remanence
measured at 10 Oe shows two distinct regimes with increasing
temperature: it decreases rapidly below 50 K and moderately above
50 K. This is consistent with the coercivity change with
increasing temperature shown in Fig. \ref{fig5}. The temperature
dependence of magnetization for superparamagnetism and
single-phase ferromagnetism can be described by their sum
\cite{PhysRevB.46.14594}:

\begin{align}
M=M_s(0)[1-(T/{T_c})^{3/2}]+n\mu[coth({\mu}H/{k_B}T)-{k_B}T/{\mu}H]
\end{align}

The first term refers to the spontaneous magnetization of the ferromagnetic contribution at temperature below the Curie temperature and the second refers
to the superparamagnetism. For the first term, $\textit{T$_c$}$ is
the Curie temperature for the ferromagnetic component, $M_S(0)$ is
the spontaneous magnetization at 0 K, and $k_B$ is the Boltzmann constant. For the second
term, \emph{n} is the density of superparamagnetic clusters and $\mu$ is
the average magnetic moment of one superparamagnetic cluster.

\begin{table}
\caption{Fitting results for samples 1E14 and 5E13 using Eq. 3.}
\begin{ruledtabular}
\begin{tabular}{ccccc}
           Samples   & $T_C$ & $M_S(0)$  & $n$   & $\mu$ \\
\hline
1E14   & 762 K & 3.6$\times$10$^{19}$ $\mu_B/g$ & 4.1$\times$10$^{14}/g$ & 8.1$\times$10$^{4}$ $\mu_B$ \\

5E13   & 750 K & 1.3$\times$10$^{19}$ $\mu_B/g$ & 5.3$\times$10$^{14}/g$ & 4.5$\times$10$^{4}$ $\mu_B$ \\

\end{tabular}
\end{ruledtabular}
\end{table}

As shown in Fig. \ref{fig5}, the temperature-dependent
magnetization is well fitted by considering both
superparamagnetism and ferromagnetism contributions. The fitting results are shown in Table I. For superparamagnetism the spontaneous magnetization can only appear below the blocking temperature. Basically for both samples, the contribution from the superparamagnetic part drops to zero slightly above 50 K. For sample 1E14, the fitting shows that one superparamagnetic cluster contains 8.1$\times$10$^4$ $\mu_B$ on average. In the
light of a theoretical calculation \citep{PhysRevLett.106.087205},
each divacancy generates 2 $\mu_B$. Assuming the magnetic moments
arising from neutral V$_{Si}$V$_C$ divacancies, one magnetic
cluster in sample 1E14 contains around 40000 V$_{Si}$V$_C$
divacancies, however, it is difficult to correlate it to a
physical cluster size. For the ferromagnetic contribution, the
best fitting gives a value for $T_c$ of around 760 K. Although the saturation magnetization
is different for samples 5E13 and 1E14 (see Fig. \ref{fig4}), both samples reveal a
similar behavior in the temperature-dependent magnetic remanence.
The ferromagnetic part of sample 5E13 is deduced to also have a
$T_c$ value around 750 K. That means the magnetic properties in both ion
irradiated SiC samples exhibit the same nature. The interplay between
the concentration of defects and the crystallinity results in a
different magnitude of magnetization. We will discuss the origin
for the coexistence of different magnetic phases in the next
section.

\begin{figure}
\includegraphics[scale=0.75]{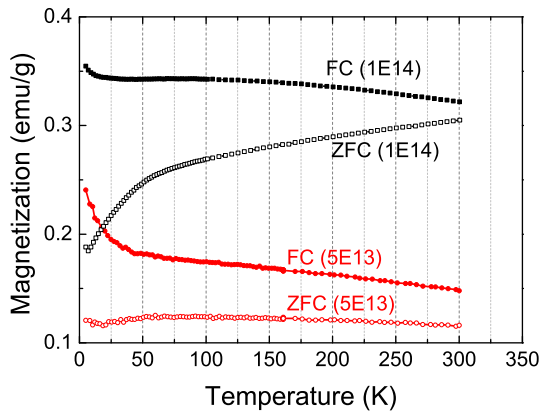}
\caption{ZFC/FC magnetization for two irradiated SiC samples with Ne ion fluences of 5$\times$$10^{13}$/cm$^2$ (5E13) and 1$\times$$10^{14}$/cm$^2$ (1E14), respectively. The thermal irreversibility has been observed for both samples. The data are shown without correcting for the diamagnetic and paramagnetic background which is negligible due to the small field.} \label{fig7}
\end{figure}

The coexistence of different magnetic phases can also be evidenced by the
zero field cooled/field cooled (ZFC/FC) measurements as shown in Fig. \ref{fig7}.
The ZFC magnetization was measured by cooling the sample from 300 K to 5 K
with zero field, then a field of 100 Oe was applied and the magnetization was measured
during warming up. The FC curves were measured by cooling the sample in a
field of 10000 Oe, then at 5 K the field was decreased to 100 Oe and the
magnetization was measured also during warming up. The ZFC/FC curves for
both samples are different from those of samples with a single magnetic
phase. For a superparamagnetic system (\emph{e.g.} magnetic nanoparticles) the ZFC
curve would increase with temperature, peak at a particular temperature
range and finally merge with the FC magnetization. For a ferromagnetic
system, when the coercivity is larger than the applied field, the ZFC
magnetization can be very small compared with the FC magnetization and
temperature-independent below the Curie temperature. When the coercivity is
small, the ZFC curves superimpose on the FC curves\cite{ISI:000239423200023}. For both SiC samples shown in Fig. \ref{fig7},
the ZFC/FC curves cannot be explained by a single magnetic phase.
They can only be understood, if assuming a superparamagnetic phase with blocking temperature around 50 K and a ferromagnetic phase with Curie temperature well above 300 K.

\subsection{Magnetic anisotropy}
\begin{figure}
\includegraphics[scale=0.75]{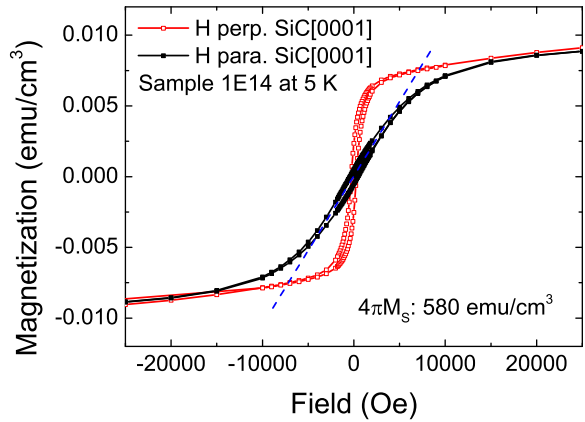}
\begin{flushleft}
\caption{Magnetization vs. field measured in-plane (field perpendicular to the SiC[0001] axis) and out-of-plane (field parallel to the SiC[0001] axis). A clear magnetic anisotropy is observed. The dashed line is used to approximately obtain the anisotropy field. The magnetization is normalized by the whole sample volume.}\label{fig8}
\end{flushleft}
\end{figure}

\textit{M-H} loops were also measured for ferromagnetic SiC with the field
both perpendicular and parallel to the sample surface. Figure \ref{fig8} shows the
comparison of the magnetization along two directions at 5 K. Note that the
diamagnetic and paramagnetic backgrounds have been subtracted for the
magnetization along both directions. Obviously, the in-plane direction is
the easy axis. It is worth noting that we do not observe any
significant difference in the M-H loops along the two in-plane directions
SiC[10\uline{1}0] and [11\uline{2}0].

Since the implanted layer is a thin layer compared with the bulk substrate (300 nm \emph{vs.} 330 $\mu$m), it is reasonable to discuss the anisotropy as a geometric effect, \emph{i.e.} as due to the shape anisotropy. We can reasonably assume a demagnetization factor 4$\pi$ for the ferromagnetic SiC\cite{coey2010magnetism}. The ferromagnetic volume fraction can thus be estimated to be 1.5$\times$10$^{-5}$, which is around 5 nm given the sample thickness of around 330 $\mu$m. The ferromagnetic thickness is much smaller than the irradiation thickness (see Fig. 1). One reason could be that the ferromagnetism actually only exists in a thin interface layer \cite{PhysRevB.72.024450,ISI:000271895500020}. Due to the dynamic annealing the vacancies in SiC can recombine with interstitial ions as well as re-distribute, finally resulting in thin planar regions which are ferromagnetic. Of course, we cannot exclude other anisotropic sources, for instance, anisotropy induced by strain.

\section{Summary and Discussion}

We have disentangled defect-induced ferromagnetism in Ne
ion irradiated 6H-SiC. The magnetization can be decomposed to three
contributions: paramagnetism, superparamagnetism and ferromagnetism. The ferromagnetism
dominates at high temperature and its Curie temperature is estimated to be
760 K.  We also observed a pronounced magnetic anisotropy with the in-plane
direction as the easy axis.

Defect-induced ferromagnetism has been observed in graphite and oxides. For
all-carbon systems, various groups have shown that the vacancies are magnetic using spin-polarized density functional theory (DFT). The moment per
vacancy is sizeable up to 1-2 $\mu_B$\cite{PhysRevLett.93.187202,PhysRevLett.99.107201}. Indeed, by scanning
tunneling microscopy experiments, Ugeda \textit{et al}. have observed a
sharp electronic resonance at the Fermi energy around a single vacancy in
graphite, which can be associated with the formation of local magnetic
moments \cite{PhysRevLett.104.096804}.  At the same time, the local environment of
the vacancy in graphite is found to strongly influence the magnetic state. Mostly,
hydrogen and nitrogen are found to play a positive role. Vacancy-nitrogen
or vacancy-hydrogen pairs result in a larger magnetic moment compared with
the vacancy alone.

For ferromagnetic SiC, a similar theoretical scenario concerning vacancies
has been adopted. First-principles calculations predict high-spin
configurations for negatively-charged Si vacancies in 3C-SiC\cite{Zhao201096}. For the case of 6H-SiC, DFT-based first-principles
calculations show that each neutral divacancy (V$_{Si}$V$_C$, an adjacent
Si and C vacancy pair) yields 2 $\mu_B$ and the tails of the defect wave functions overlap, resulting in
ferromagnetic coupling between adjacent divacanices\cite{PhysRevLett.106.087205}. Indeed, such V$_{Si}
$V$_C$ divacancies are observed by positron annihilation spectroscopy in
6H-SiC after neutron and Ne ion irradiation\cite{Li201198,PhysRevLett.106.087205}. Considering the
magnetic properties of defective SiC presented in this work, we can
envision the following picture: (1) the defects in SiC are non-homogenously
distributed, (2) the isolated, uncoupled defects contribute to the
paramagnetism and (3) the coupled defects result in the superparamagnetic
and ferromagnetic components.

Now we discuss the simultaneous appearance of ferromagnetic and
superparamagnetic components in ion irradiated SiC. In ion irradiated
graphite, a paramagnetic component corresponding to independent magnetic
moments and a ferromagnetic contribution are observed\cite{PhysRevB.76.161403,PhysRevB.81.214404}. The ferromagnetic component shows a linear temperature dependence and can be
well explained by a two-dimensional Heisenberg model. Recently by careful
angular dependent near-edge x-ray-absorption fine-structure analysis (NEXAFS), He et
al., observed defect electronic states near the Fermi level, which are
extended in the graphite basal plane\cite{PhysRevB.85.144406}. In chemically processed
graphite, multilevel ferromagnetism was observed and is attributed to the
weak coupling between ferromagnetic regions and to the ferromagnetism
inside the defect regions\cite{PhysRevB.71.100404}. Therefore, in ion irradiated SiC the
ferromagnetic coupling is expected to first arise within regions with proper vacancy
concentrations. Some regions are connected to be big enough and behave as a
ferromagnet. Some regions with ferromagnetically coupled defects are small and behave as
superparamagnets. However, the ferromagnetic and superparamagnetic
components in the two samples irradiated with different fluences 5E13 and
1E14 (Fig. \ref{fig6} and Fig. \ref{fig7}) are very similar regarding the Curie temperature and
the superparamagnetic blocking temperatures. This fact essentially
excludes a depth distribution of defects as the origin of the magnetic
inhomogeneity. In a recent work on Gd-doped GaN using large-scale first-principles electronic structure calculations\cite{PhysRevB.86.180401}, Thiess \emph{et al.} find that intrinsically the Ga
vacancies tend to cluster, resulting in the co-existence of ferromagnetism
and superparamagnetism, which could also explain the present experimental
results in our SiC.

As a conclusion, we have observed a magnetic inhomogeneity in SiC
irradiated by Ne ions. Three magnetic phases (paramagnetic,
superparamagnetic and ferromagnetic phases) can be identified by numerical
fitting. Moreover, the ferromagnetic SiC samples exhibit a pronounced
magnetic anisotropy with the easy axis along the SiC(0001) basal plane. By
comparing with the model and experimental data in literature, we
tentatively attribute the magnetic properties to the inhomogenous
distribution of defects in SiC and the preferential clustering along the
basal plane.

\section{Acknowledgement}

The work is financially supported by the Helmholtz-Gemeinschaft Deutscher
Forschungszentren (VH-NG-713 and and VH-VI-442). Y. Wang thanks the China Scholarship
Council for supporting his stay at HZDR. The authors also acknowledge the partial support by the International Science and Technology Cooperation Program of China (2012DFA51430).


\end{document}